\documentclass[twocolumn,showpacs,secnumarabic,amssymb,nobibnotes,aps,pre]{revtex4}
\usepackage{amsmath}
\usepackage{latexsym}
\usepackage{color}
\usepackage{graphicx}
\usepackage{textcomp}
\usepackage{hyperref}

\begin{document}

\title{Disentangling Growth and Decay of Domains During Phase Ordering}
\author{Suman Majumder}\email[]{smajumder.@amity.edu, suman.jdv@gmail.com}
\affiliation{Amity Institute of Applied Sciences, Amity University Uttar Pradesh, Noida 201313,
India
}




\date{\today}

\begin{abstract}
Using Monte Carlo simulations we study the phase ordering dynamics of a \textit{multi}-species system modeled via the prototype $q$-state Potts model. In such a \textit{multi}-species system, we identify a spin states or species as the \textit{winner} if it has survived as the majority in the final state, otherwise we mark them as \textit{loser}. We disentangle the time ($t$) dependence of the domain length of the \textit{winner} from \textit{losers}, rather than monitoring the average domain length obtained by treating all spin states or species  alike. The kinetics of domain growth of the \textit{winner} at a finite temperature in space dimension $d=2$ reveal that the expected Lifshitz-Cahn-Allen scaling law $\sim t^{1/2}$ can be observed with no early-time corrections, even for system sizes much smaller than what is traditionally used. Up to a certain period, all the others species, i.e., the \textit{losers}, also show a growth that, however, is dependent on the total number of species, and slower than the expected  $\sim t^{1/2}$ growth.  Afterwards, the domains of the \textit{losers} start decaying with time, for which our data strongly suggest the behavior $\sim t^{-z}$, where $z=2$ is the dynamical exponent for nonconserved dynamics. We also demonstrate that this new approach of looking into the kinetics also provides new insights for the special case of phase ordering at zero temperature, both in $d=2$ and $d=3$.

\end{abstract}


\maketitle

\section{INTRODUCTION}
Existence of a discrete number themodynamically degenerate ordered phases at low-temperature is the characteristics of many systems, e.g., ordering alloys, adsorbed atoms and rare gases on surface, intercalation compounds \cite{moncton1977,berker1978,bak1979theory}. Theoretically and computationally such systems can effectively be understood via a \textit{multi}-species model. In this context, the prototype is the $q$-state Potts 
model \cite{Potts_RMP}. It has been used extensively, especially, to investigate phase transitions associated with such systems. Slight variation of the model makes it useful in diverse applications, be it in biological cell sorting or in modeling active matter system \cite{graner1992,chen2007,szabo2013,guisoni2018,chatterjee2020,mangeat2020}. In the context of phase transition,  among all the studied problems one is particularly intriguing -- to understand the nonequilibrium kinetics in reaching the final equilibrium state following a quench from an initial disorder state to a temperature below the transition temperature, where the equilibrium is an ordered state. Such a process goes by the name of phase ordering or coarsening. Its dynamics is highlighted by the formation and growth of domains of like-species \cite{Bray2002,Puri_book}. The phenomenon is characterized by the scaling of various morphology-characterizing functions, viz., the two-point equal-time correlation function exhibits a scaling of the form  
\begin{eqnarray}\label{scld_Crt}
C(r,t)&=& \tilde{C} [r/\ell(t)]\label{scaling_of_Cr},
\end{eqnarray}
where $\tilde{C}$ is a time ($t$)-independent master function, and $\ell(t)$ is the average linear domain size which is expected to grow as 
\begin{eqnarray}\label{LCA_law}
\ell(t) \sim t^{\alpha}, 
\end{eqnarray}
with an exponent $\alpha=1/2$, refer to as the Lifshitz-Cahn-Allen (LCA) law \cite{lifshitz1962,allen1979}. It turns out that $\alpha=1/z$ where $z$ is the equilbrium dynamical exponent \cite{Bray2002}. 
\par
Most of the simulation studies on phase-ordering dynamics of the Potts model were focused on estimating the value of $\alpha$ for $q \ge 2$. Early results for smaller system sizes at temperature $T>0$ where there is no pinned or frozen dynamics, showed that the exponent decreases from $\alpha=1/2$ as $q$ increases from $q=2$ \cite{safran1981,sahni1983}. Later, using results from relatively larger system sizes for $q$ as larges as $q=64$ it was shown that $\alpha=1/2$ can only be realized in the long-time limit at subcritical temperatures \cite{sahni1983_2,kaski1985,ferrero2007}. 
However, for square lattices the problem of presence of pinned configurations exists at low but finite temperatures \cite{ferrero2007,ibanez2007}. On the other hand, due to the presence of pinned or frozen dynamics  for $q \ge d+1$, zero-temperature phase ordering is always of special interest both in $d=2$ and $3$ \cite{lifshitz1962,safran1981,loureiro2010,olejarz2013}. There, the domain growth is found to be significantly slower than the expected LCA law. For that matter, arguments in favor of logarithmic growth were also proposed \cite{safran1981}. In $d=3$, even for the $q=2$ Potts model which is basically the spin $1/2$ Ising model, the issue of domain growth at $T=0$ is not well settled \cite{lippiello2008,olejarz2011,olejarz2011zero,chakraborty2016,das2017,vadakkayil2019}. Simulation studies of relatively larger system sizes even suggested a possible crossover in the domain growth from an early time $\sim t^{1/3}$ behavior to the asymptotic $\sim t^{1/2}$ behavior \cite{lippiello2008,vadakkayil2019}.
\par
It must have been quite clear by now that in the sutdy of phase ordering the primary quantity which one measures is the average linear size of the domains $\ell(t)$ as a function of time. To calculate $\ell(t)$, scaling property of any morphology-characterizing function is used, viz., from the decay of the two-point equal-time correlation function, from the first moment of the structure factor, and from the first moment of the domain-length distribution function \cite{Bray2002}. The measured $\ell(t)$ from different methods are proportional to each other. At this point it is worth noting that in phase ordering the dynamics is nonconserved, i.e., the global order parameter, viz., the magnetization, does not remain fixed to its starting value during the entire evolution of the system. In other words, even though one typically begins with a uniform mixture having equal number of every possible spin states or species, eventually ends up with one of the spin states or  species emerging as the majority or \textit{winner} in the final state. All other spin states or species which we refer to as \textit{losers}, are present in negligible amount. So, it is easy to perceive that among all the spin states or species, domains of only the \textit{winner} will always be growing before reaching the finite-size limit. Alongside domains of \textit{losers} must be declining after a certain time and almost vanishing in the final state. When one calculates the average domain length $\ell(t)$, this distinction of the spin states or species in terms of \textit{winner} or \textit{losers} is not taken into account. Thus, the effect of decay of the \textit{losers} is always embedded in the measured $\ell(t)$. Nevertheless, till date no attempts have been made to disentangled the kinetics of the \textit{winner} from \textit{losers}.  
\par
Here, keeping the above issues in mind, by means of Monte Carlo (MC) simulations we study the phase ordering dynamics of the $q$-state Potts model with a focus on monitoring the kinetics of individual spin states or species by categorizing them as the \textit{winner} or \textit{losers}. Our results for $2\le q\le 10$ at a finite temperature for domain growth of the \textit{winner} show no early-time correction to the expected scaling law. Apart from the growth of the \textit{winner}, the kinetics of the \textit{losers} exhibit a non-monotonic behavior with a long-time power-law decay $\sim t^{-2}$, independent of the total number of spin states or species $q$. We also demonstrate that our approach of monitoring the kinetics of the \textit{winner} provides improved results for the special case of phase ordering of the $q$-state Potts model at $T=0$ in space dimension $d=2$. Furthermore in $d=3$, for $q=2$, that corresponds to the Ising model, we show that all the features that were expected to be observed in large systems can essentially be captured in relatively much smaller systems. 
\par
The remaining of the paper is organized in the following way. In Section\ \ref{methods}, we describe the model and details of the performed simulations. Then we present results in Sec.\ \ref{results}, where as we go along we also describe the calculations of different observables. At the end in Sec.\ \ref{conclusion}, we provide a brief summary, conclusion and outlook.

\section{Model and Method}\label{methods}
The Hamiltonian of the $q$-state Potts model is given by
\begin{eqnarray}\label{Potts}
 H = -J\sum_{\langle ij \rangle}\delta_{\sigma_i, \sigma_j};~\sigma_i=1,2,\dots, q;~ J>0, 
\end{eqnarray}
where $\sigma_i$ is the spin state or species type $\sigma$ of the $i$-th site, and the summation is over all possible pairs of nearest neighbor indicated by $\langle ij \rangle$. The model exhibits an order-disordered transition as a function of temperature. 
For a square lattice, the corresponding transition temperature is $T_c= J/k_B\ln(1+\sqrt{q})$, 
where $k_B$ is the Boltzmann constant \cite{Potts_RMP}. For $q\le 4$ the transitions are second order whereas for $q>5$ they are first order. 
\begin{figure*}[t!]
\includegraphics*[width=0.48 \textwidth]{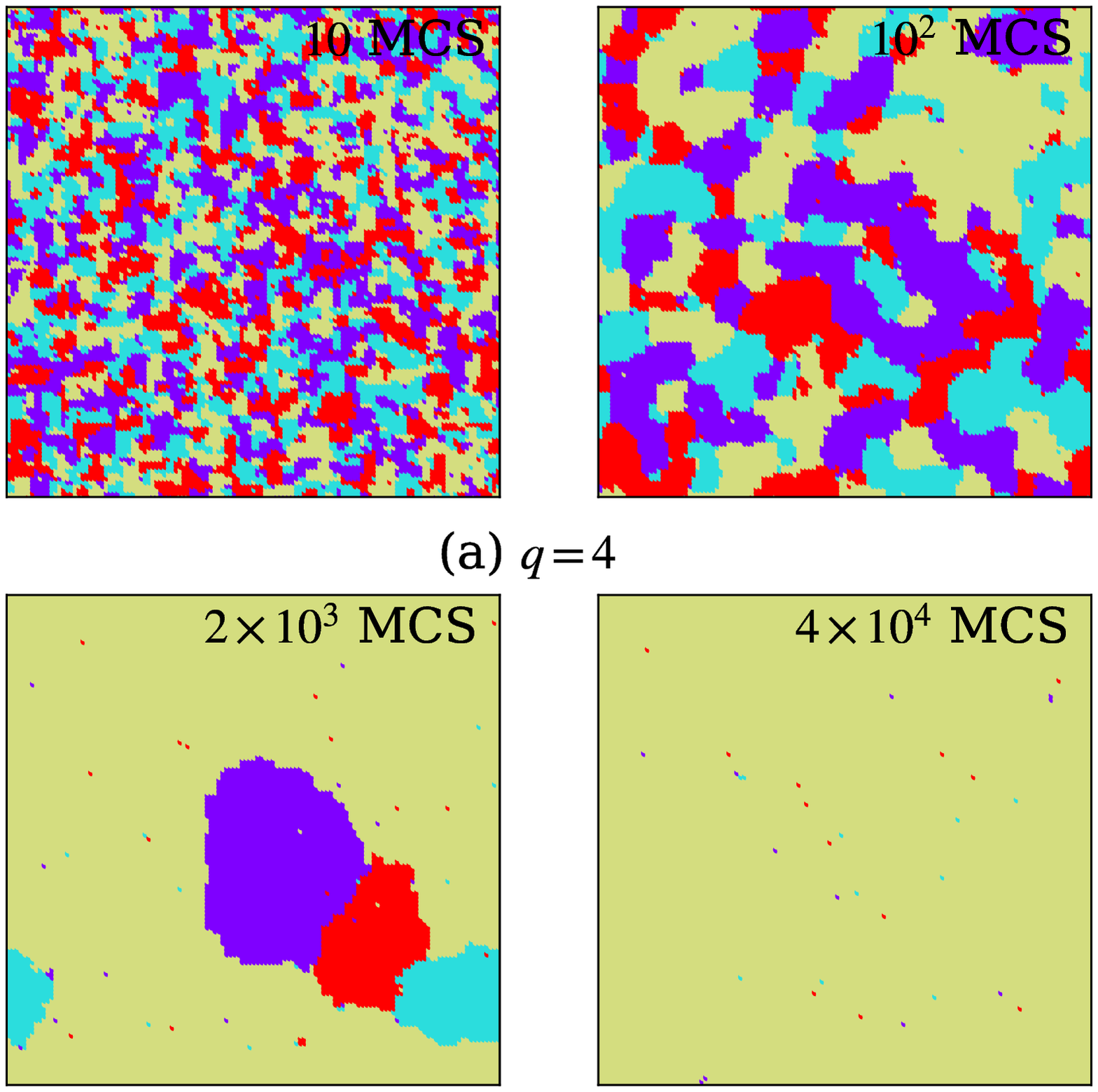}~~~~
\includegraphics*[width=0.48\textwidth]{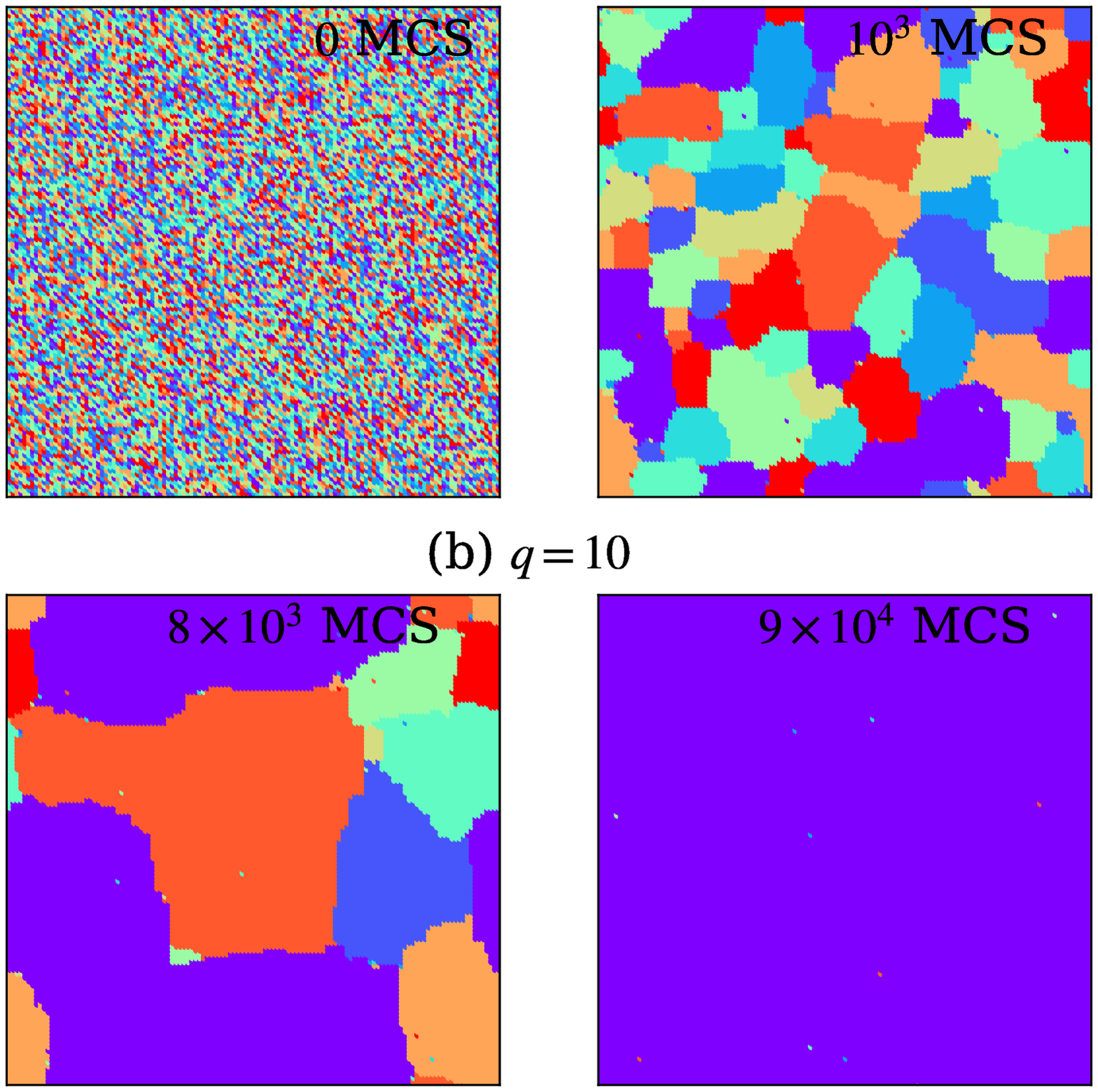}
\caption{\label{snapT0.6} Time evolution snapshots depicting the general sequence of events during phase-ordering dynamics of the $q$-state Potts model , at $T=0.6T_c$ in $d=2$ using a square lattice with linear dimension $L=128$, for two different choices of $q$, as indicated. Different colors represent different spin states or species.}
\end{figure*}
\par
We study the model \eqref{Potts} by performing MC simulations on square and cubic lattices, respectively, for $d=2$ and $3$, having linear dimensions $L$ along each Cartesian direction. We apply periodic boundary conditions in all possible directions. Dynamics in our simulations is introduced via the standard Glauber spin-flip mechanism that changes the global concentration of each spin state or species during the evolution, and finally ends up with one of the spin states or species as the majority \cite{newman_book,landau_book}.  A typical MC move consists of randomly choosing a site having the state $\sigma_i$ followed by changing its state to any of the remaining $q-1$ states. The move is then accepted in accordance with the standard Metropolis algorithm with a probability
\begin{equation}
 p_i=\textrm{min}[1,\exp(-\Delta E/k_BT)],
\end{equation}
 where $\Delta E$ is the change in energy due to the attempted move. $L^d$ such attempts form a MC sweep (MCS) setting the unit of time.

\par
To start our simulations, as initial condition we prepare a random mixture containing all possible ($q$) spin states or species in equal proportion, mimicking a  
high-temperature ($T=\infty$) paramagnetic phase. 
Such configurations are then suddenly quenched below the critical temperature $T_c$ by fixing $T=0.6T_c$. In our simulations, 
$J/k_B$ sets the unit of temperature, and for convenience $J$ and $k_B$ are set to unity. 
We also perform simulations at $T=0$ where the criterion for the Metropolis update of the system is slightly different. There, an attempted move is accepted if $\Delta E \le 0$. All the results presented are averaged over more than $100$ independent realization accounting for different starting configurations and thermal noise during evolution.

\section{Results}\label{results}
The results are presented in two subsections. In the first subsection, we discuss the results for $d=2$ at the finite temperature $T=0.6T_c$. In the other one, we present results for $T=0$ in $d=2$ and $3$.

\subsection{Phase Ordering at finite $T$ in $d=2$  }\label{finite_T}
When a disordered system containing homogeneous mixture of $q$ spin states or species is quenched below $T_c$, it becomes unstable to fluctuations and starts evolving towards an ordered state where only one of the spin states or species survives as the majority. Snapshots for such  an evolution using a square lattice at $T=0.6T_c$ is presented in Fig.\ \ref{snapT0.6} for $q=4$ and $10$. In both cases the process starts with formation of domains of like spins as represented by the snapshots at $t=10$ MCS. These domains then start growing with time, as evident from the snapshots at $t=10^2$ MCS and $t=10^3$ MCS, respectively, for $q=4$ and $10$. Afterwards, once the domains of individual spin state or species have attained a considerable size, one of the species keeps on growing further while the others start decaying and eventually almost vanish. This leads to the emergence of the \textit{winner} species which stays as the majority in the final configuration. For example, in the case presented in Fig.\ \ref{snapT0.6}, the yellow and purple states are \textit{winners}, respectively, for $q=4$ and $10$. There all other states are marked as \textit{losers}. By arranging the spin states or species in descending order in terms of the total number of sites in the final configuration having that particular spin state or species, we assign a ranking $\Re_{\sigma}$ to them where 
\begin{equation}
 \Re_{\sigma} \in \{1,2,\dots q\}.
\end{equation}
The \textit{winner} will have $\Re_{\sigma}=1$ and all other spin states or species with $\Re_{\sigma} > 1$ correspond to \textit{losers}. $\Re_{\sigma}=q$ indicates that the number of sites in the final configuration belonging to that spin state or species is the least. Note that the condition $\Re_{\sigma}=\sigma$ is not necessarily true.
\par
We start with measuring the primary quantity of interest, the growing characteristic length scale, which in the present case is the linear size of the domains. However, in contrast to the usual practice, we measure the length of the domains as a function time of each spin state or species and store them during the simulation. The domain of an individual spin state or species is measured by mapping the Potts spins of every lattice site $i$ to an Ising-like scalar variable as \cite{das_Potts,majumder_Potts}
\begin{equation}\label{Ising-like}
\psi_i^n=2\delta_{\sigma_i,n}-1;~n=1,2,\dots q.
\end{equation}
This gives a lattice of spin $\pm 1$ for every value of the Potts spin $n$. By scanning through these lattices in $x$, $y$, and $z$ (for $d=3$) directions, we measure the total number of interfaces or defects  $N_{\rm def}$ in each of the constituent Ising chains of the lattice, of course, by taking into account of the periodic boundary conditions. For each Ising chain of length $L$, the total number of linear domains will be $N_{\rm ld}=N_{\rm def}/2$, except for the special cases where all the spins in the chain are either $+1$ or $-1$. In the former case, the number of linear domains will be $1$ and for the latter there will be no domains of $+1$ spins. Finally, by using the total number $N_{+}$ of $+1$ spins present in the lattice, we calculate the average linear size of domains of the spin state or species $n$ as 
\begin{equation}
 \ell_{n}=\langle N_+/N_{\rm ld} \rangle,
\end{equation}
where the bracket indicates averaging over all possible directions. Thus, by using the corresponding Ising-like lattice for every spin state or species, we calculate the average linear size of domains of every spin state or species. Finally, we do a change of identity of the domain lengths by $\ell_{n} \rightarrow \ell_{\Re_{n}}$ using the previously introduced ranking of the spin states or species. The average of domain lengths obtained from different trajectories are then calculated in terms of $\ell_{\Re_{n}}$. Thus, from now onward $\ell_1$ corresponds to the average linear domain size of the \textit{winner} and $\ell_{\Re_{n}}$ with $\Re_{n} > 1$ correspond to \textit{losers}. In the traditional way, different spin states or species are not distinguished, and  average domain length $\ell$ is calculated by considering them alike, so that 
\begin{equation}\label{lt_avg}
 \ell = \frac{1}{q}\sum_{\Re_{n}=1}^{q}\ell_{\Re_{n}}.
\end{equation}
\begin{figure}[t!]
\centering
\includegraphics*[width=0.48\textwidth]{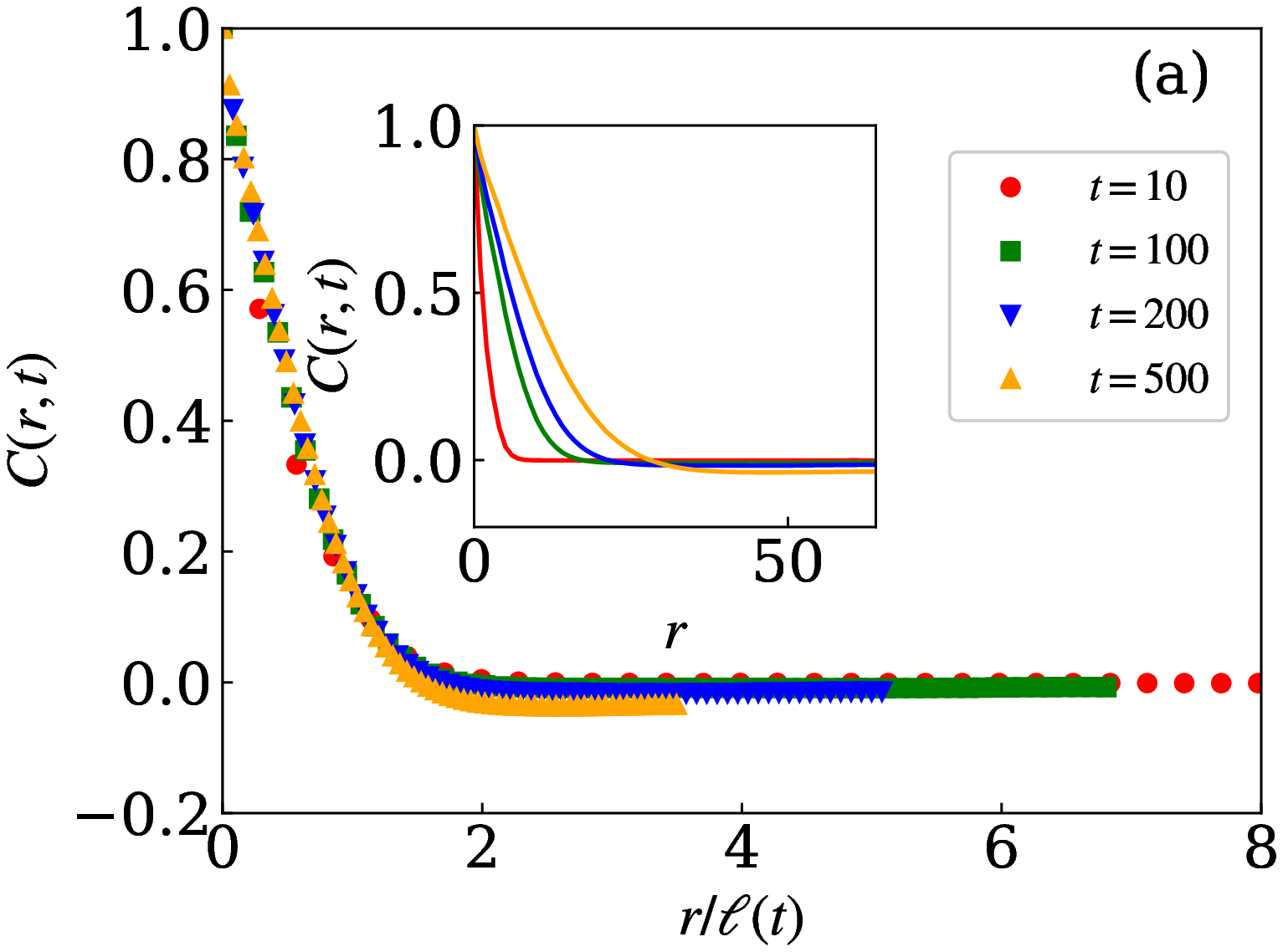}\\
\includegraphics*[width=0.48\textwidth]{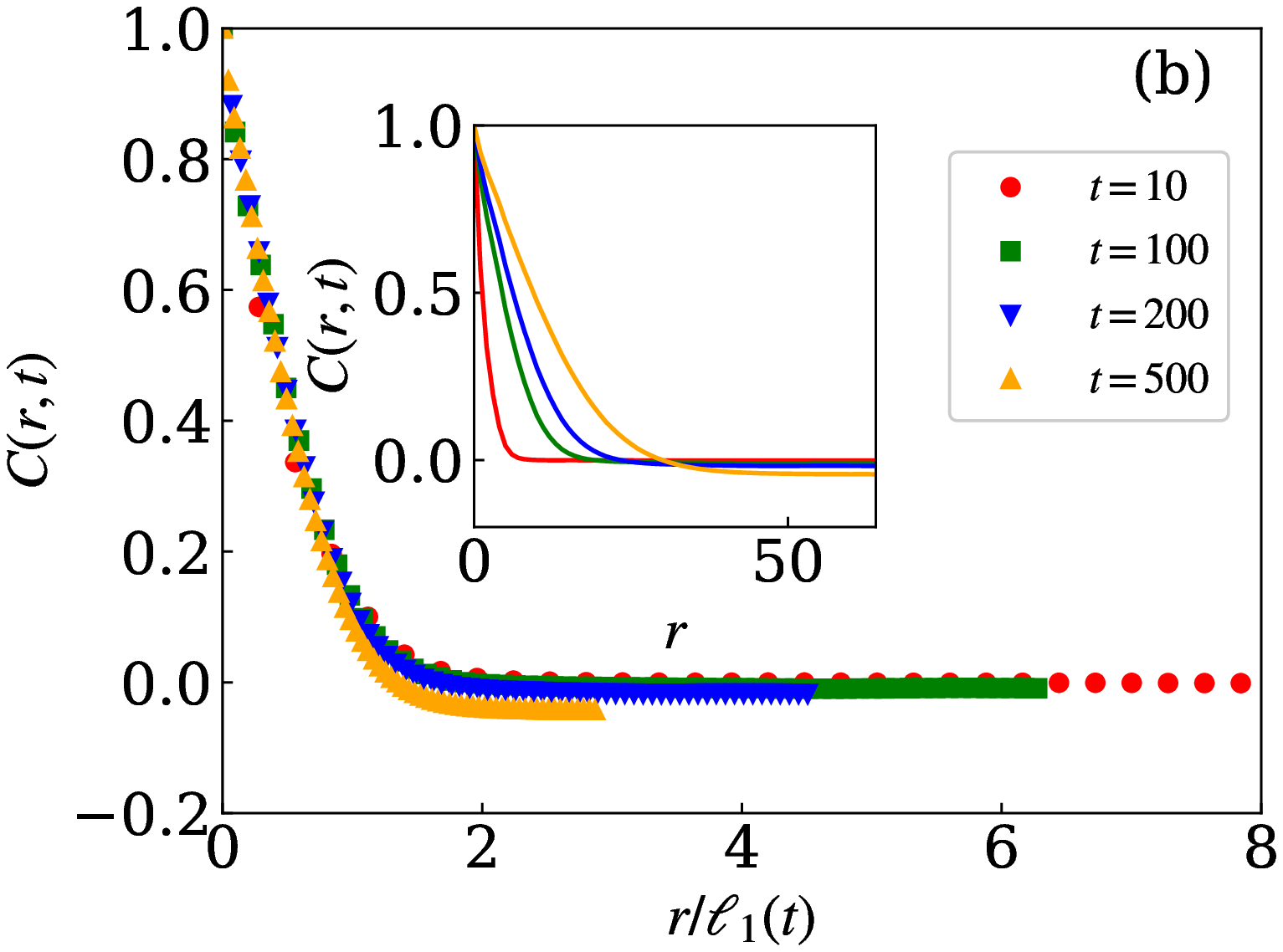}
\caption{\label{corr_q4_T0.6} Demonstration of fulfillment of the scaling criterion \eqref{scld_Crt} of the two-point equal-time correlation function during phase ordering in a square lattice with $L=128$ for $q=4$, at $T=0.6T_c$. In (a) the average linear domain length $\ell(t)$ is used as the scaling factor for the correlation function $C_{\rm avg}(r,t)$, obtained from averaging over all the spin states or species. The same for the correlation function of the \textit{winner} species $C_1(r,t)$, using the domain length of the \textit{winner} $\ell_1(t)$ is shown in (b). The insets show the corresponding unscaled data at different times.}
\end{figure}
\begin{figure}[t!]
\centering
\includegraphics*[width=0.48 \textwidth]{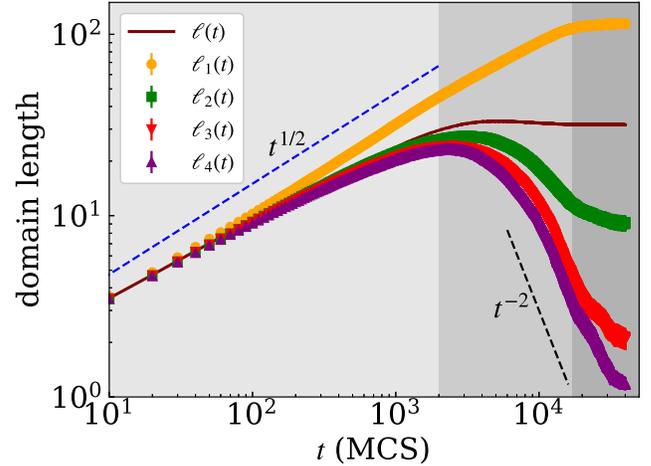}
\caption{\label{length_q4_T0.6} Time dependence of different domain lengths measured for a phase-ordering system with $q=4$ using a square lattice having $L=128$, at $T=0.6T_c$. Here, $\ell(t)$ correspond to the traditionally calculated domain length obtained by averaging over all spin states or species. The other lengths presented are for different spin states or species in terms of their  ranking using $\Re_{n}$ (see text for details). According to the ranking, $\ell_1$ corresponds to the domain length of the \textit{winner} species and the rest are all \textit{losers}. The dashed blue line represents the expected LCA growth $\sim t^{1/2}$. The dashed black line represents a power-law decay $\sim t^{-2}$. The shades in grey scale are introduced to roughly mark different regimes.}
\end{figure}
\par
Before proceeding further with the domain length for the \textit{winner} and \textit{loser}, we verify whether $\ell_1$ satisfies the criterion of a scaling phenomenon \eqref{scld_Crt}. For that we  calculate the 
two-point equal-time correlation function averaged over all the spin states or species as 
\begin{eqnarray}\label{corr_func}
 C_{\rm avg}(r,t) =\frac{1}{q} \sum_{n=1}^{q}\left[ \langle \psi_i^n(t)\psi_j^n(t) \rangle 
 -\langle \psi_i^n(t) \rangle \langle \psi_j^n(t) \rangle \right],
\end{eqnarray}
where $r$ is the scalar distance between any two sites $i$ and $j$. The corresponding scaling plots of $C_{\rm avg}(r,t)$ by using $\ell(t)$ is presented in Fig.\ \ref{corr_q4_T0.6}(a) for phase ordering in a square lattice system with $q=4$ at $T=0.6T_c$. The collapse of data for different times implies the fulfillment of the scaling criterion \eqref{scld_Crt}. Next, we calculate two-point equal-time correlation function of the winner species 
\begin{eqnarray}\label{winner_corr_func}
 C_1(r,t) = \langle \phi_i^{1}(t)\phi_j^{1}(t) \rangle 
 -\langle \phi_i^{1}(t) \rangle \langle \phi_j^{1}(t) \rangle,
\end{eqnarray}
where $\phi_i^{1}$ is the Ising-like lattice variable obtained using the \textit{winner} spin state or species. The corresponding scaling plot of $C_1(r,t)$ using the domain length of the \textit{winner} $\ell_1(t)$ as the scaling factor, is shown in Fig.\ \ref{corr_q4_T0.6}(b). Reasonably good collapse of data confirms the satisfaction of the required scaling criterion \eqref{scld_Crt}. In fact at early times, for small $r$ the collapse of data in (b) is superior to the data presented in (a).
\begin{figure}[t!]
\includegraphics*[width=0.48 \textwidth]{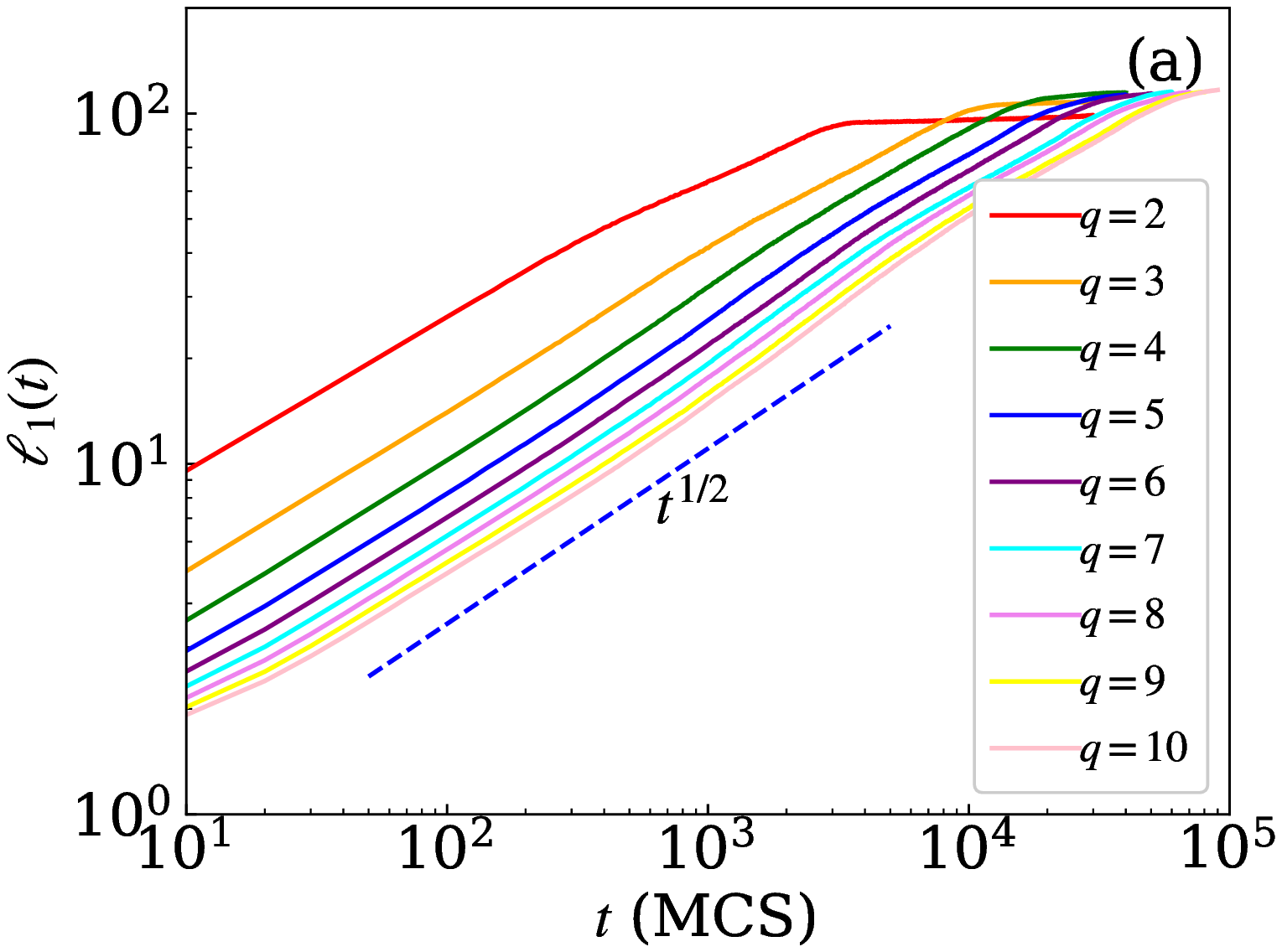}\\
\includegraphics*[width=0.48 \textwidth]{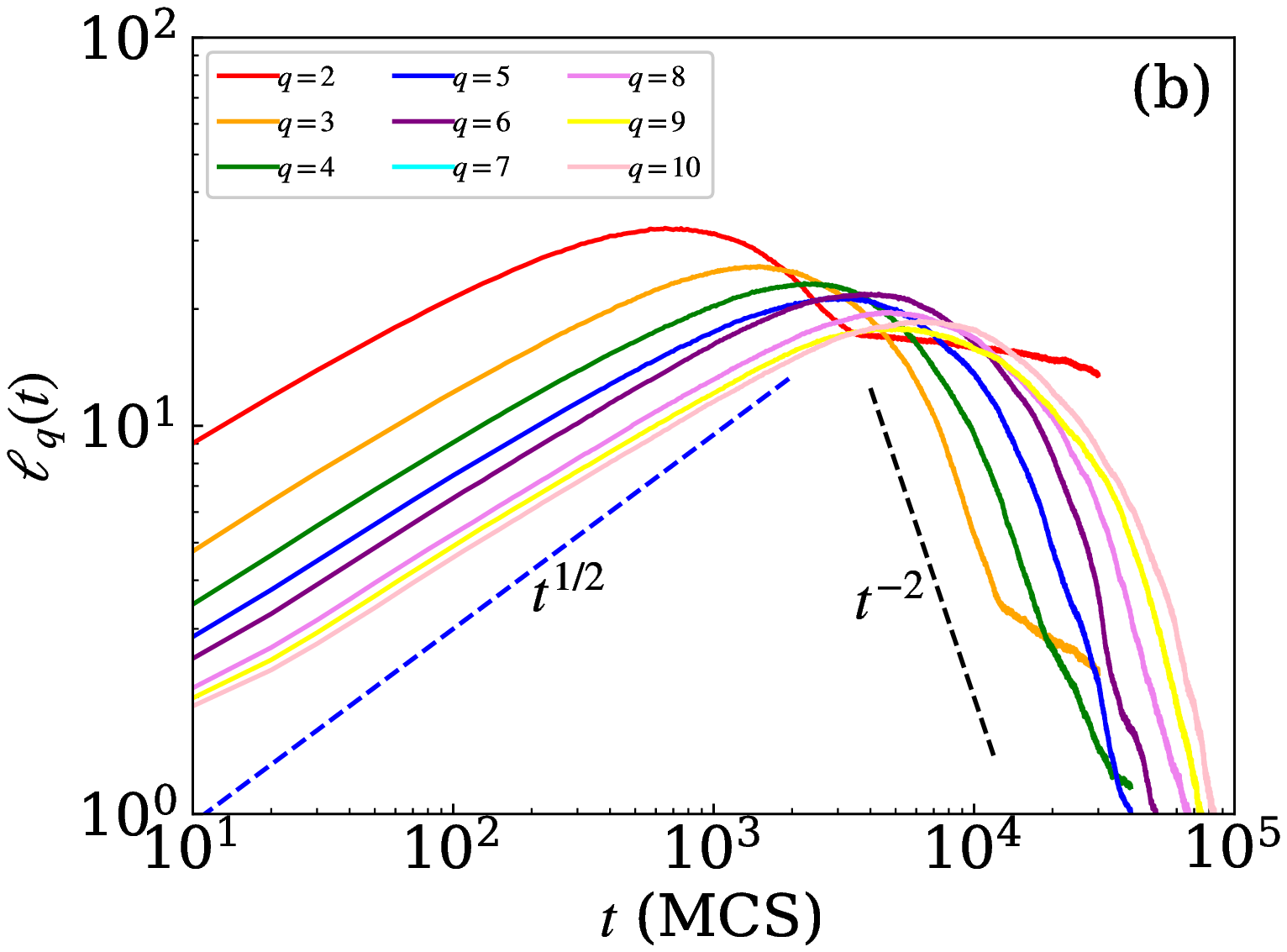}
\caption{\label{length_diffq_T0.6} Time dependence of the domain length of (a) the \textit{winner} $\ell_1$ and (b) \textit{loser} $\ell_q$ for different values of $q$ at $T=0.6T_c$ using a square lattice system of size $L=128$. The blue dashed lines represent the LCA growth law $\sim t^{1/2}$, and the black dashed lines represent the power-law decay $\sim t^{-2}$. }
\end{figure}
\par 
Having the fulfillment of a scaling phenomenon established, we move on to deal with time dependence of domain lengths in Fig.\ \ref{length_q4_T0.6}. There, the data for $\ell(t)$, obtained by considering all $q$ spin states or species, show a much slower growth compared to the expected LCA law $\sim t^{1/2}$. Such an inconsistency with the expected growth law has always been a problem for smaller system sizes, like $L=128$, presented here. This could be explained via a combination of finite-time corrections and early advent of finite-size effects. In such cases, there have been attempts to extract the asymptotic growth by applying finite-size scaling via an ansatz considering an initial domain length \cite{Majumder2010,Majumder2011,das2012finite}. However, the trend has always been to use larger system sizes to access larger length and time scales. As mentioned, we extract the time dependence of $\ell_{\Re_{n}}$ by categorizing the spin states or species via a ranking. The different domain lengths obtained this way are also plotted in Fig.\ \ref{length_q4_T0.6}.  
There, the data for domain length $\ell_1(t)$ of the \textit{winner} show remarkable consistency with the expected $\sim t^{1/2}$ growth over an extended period starting from very early time. Followed by that it enters a regime where it still grows, however, with a smaller exponent. Finally, it hits the finite-size limit and remains constant as long as we ran the simulations. The intermediate regime is essentially an extended crossover regime from the scaling regime to the finite-size limit. In contrast, the data for $\ell(t)$ show a much sharper crossover to the finite-size limit.

\par
Intriguing is the behavior of the domain length of the \textit{losers} as shown in Fig.\ \ref{length_q4_T0.6} by the data for $\ell_{\Re_{n}}$ with $\Re_{n}=2,3, \dots q$. In the scaling regime (lightest grey zone) they show a much slower growth compared to the \textit{winner} or the expected LCA law. The slower growth in this regime is the root cause of the slower growth for $\ell(t)$ in the scaling regime. After the scaling regime, the growth of the \textit{losers} gets seized for a very brief period, and   eventually it enters into a decay regime that coincides with intermediate crossover regime for $\ell_1(t)$. For $\ell_q$, i.e., $\ell_4$, the domain length of the last ranked spin state or species, the decay seems to be consistent with $\sim t^{-2}$ indicating a $\sim t^{-z}$ behavior, of course, subjected to further verification for other values of $q$, as will be done subsequently.  
\begin{figure}[t!]
\centering
\includegraphics*[width=0.48 \textwidth]{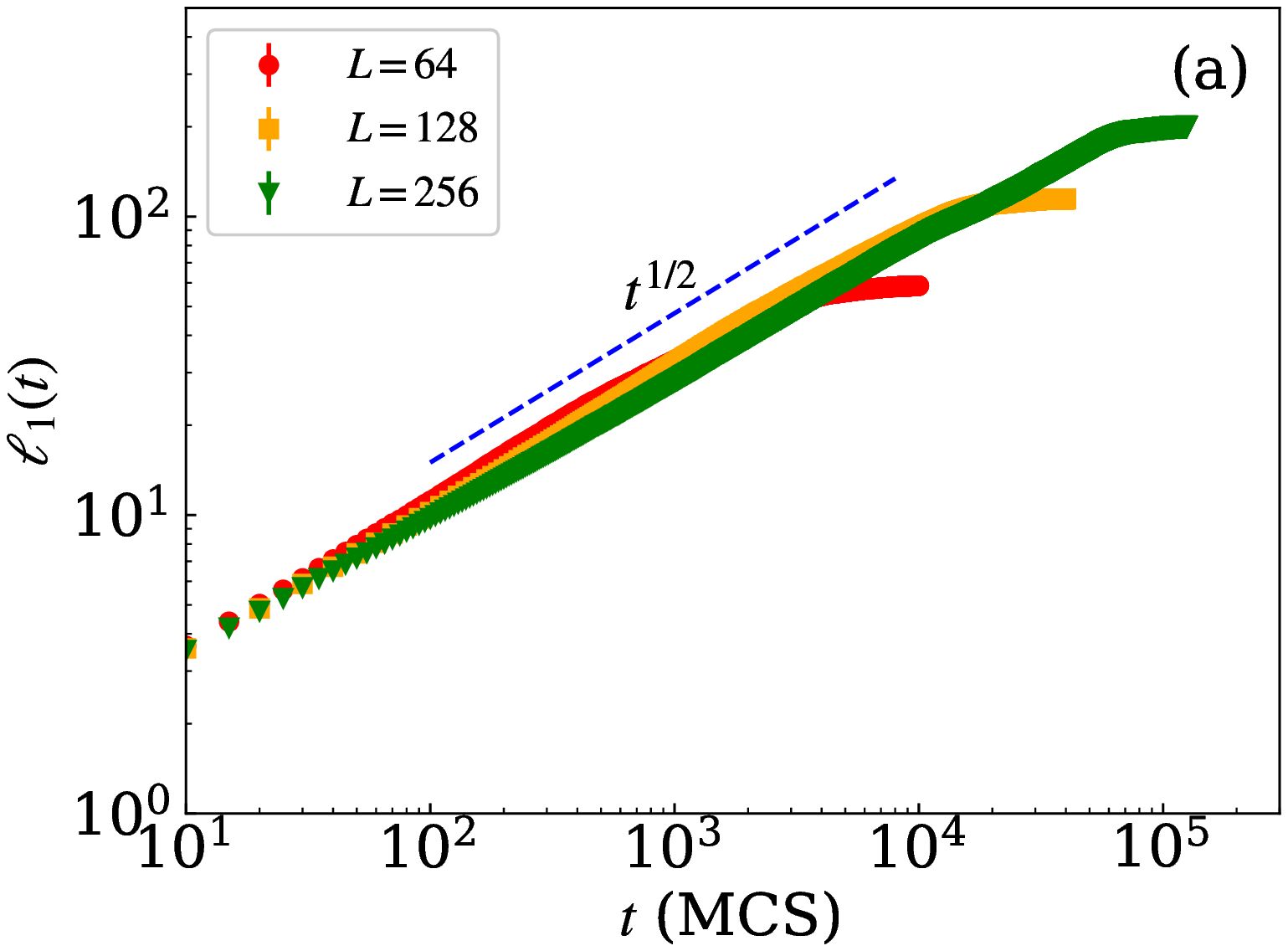}\\
\includegraphics*[width=0.48 \textwidth]{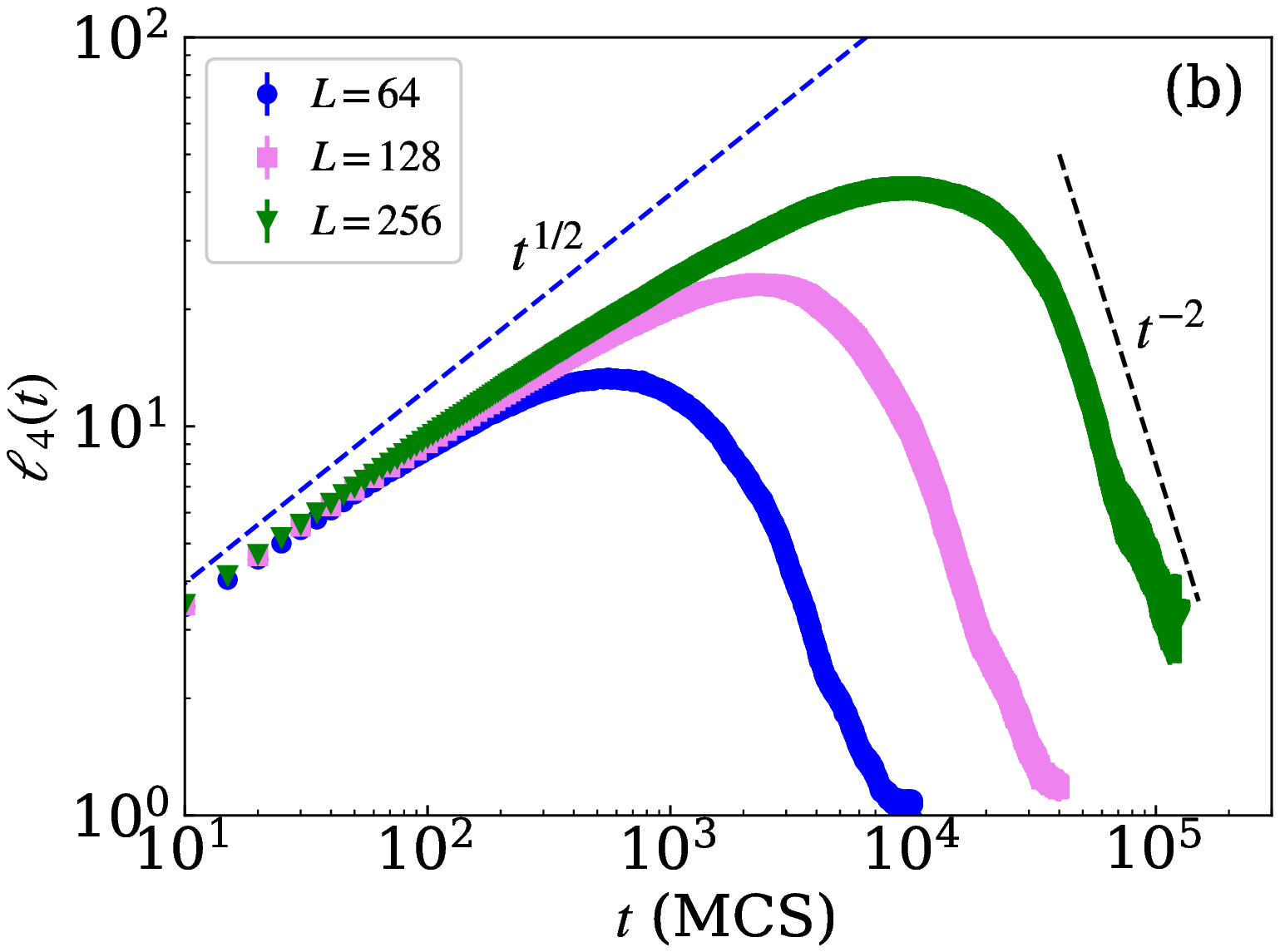}
\caption{\label{length_diffL_T0.6} System size dependence of the kinetics of domain length of (a) the \textit{winner} and (b) \textit{loser} for $q=4$ at $T=0.6T_c$. Square lattice system with three different $L$, as mentioned, are used. The dashed blue lines in (a) and (b) represent the LCA growth. The dashed black line in (b) represents a power-law decay $\sim t^{-2}$.}
\end{figure}
\begin{figure*}[t!]
\centering
\includegraphics*[width=0.48 \textwidth]{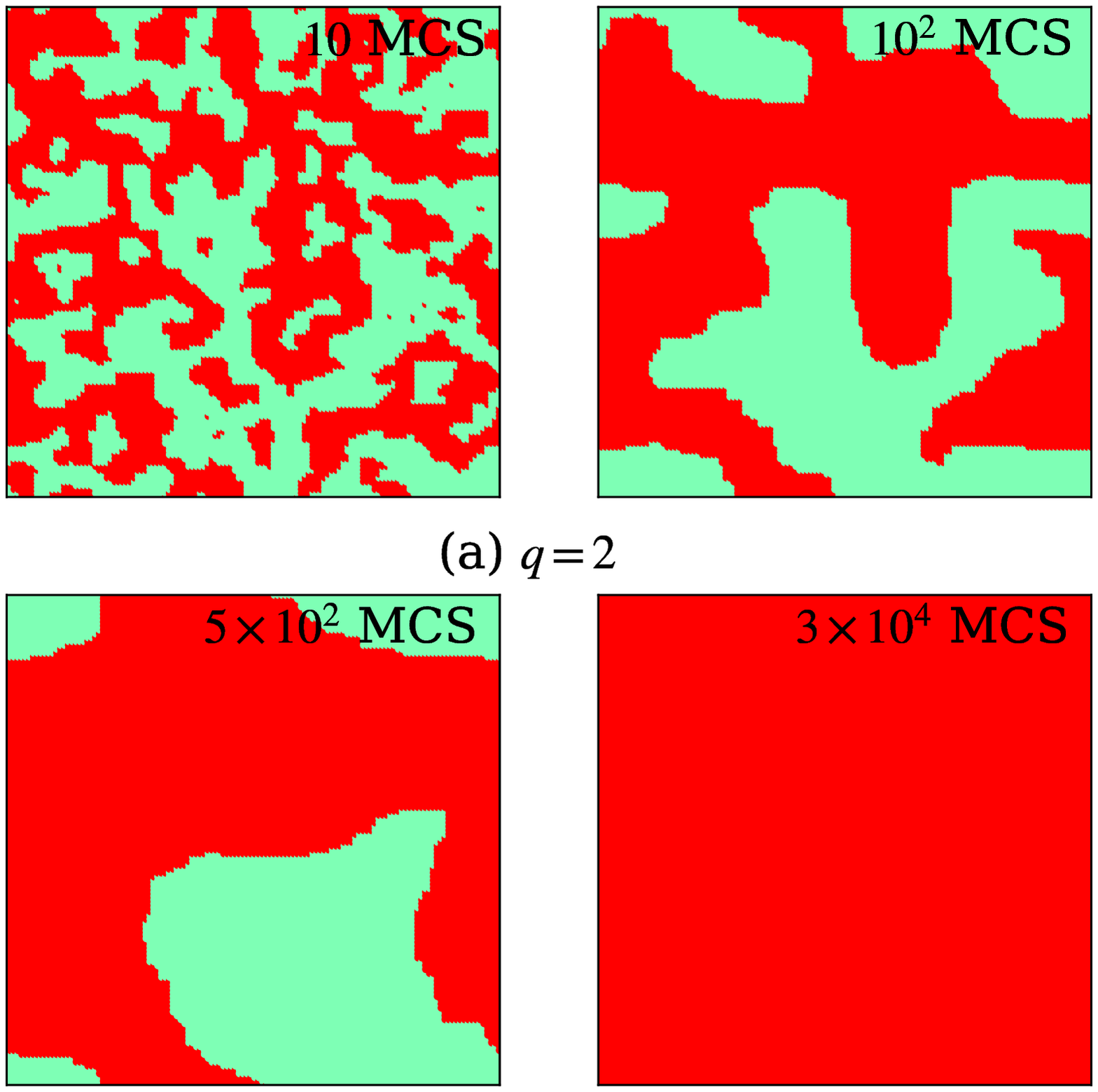}
~~
\includegraphics*[width=0.48 \textwidth]{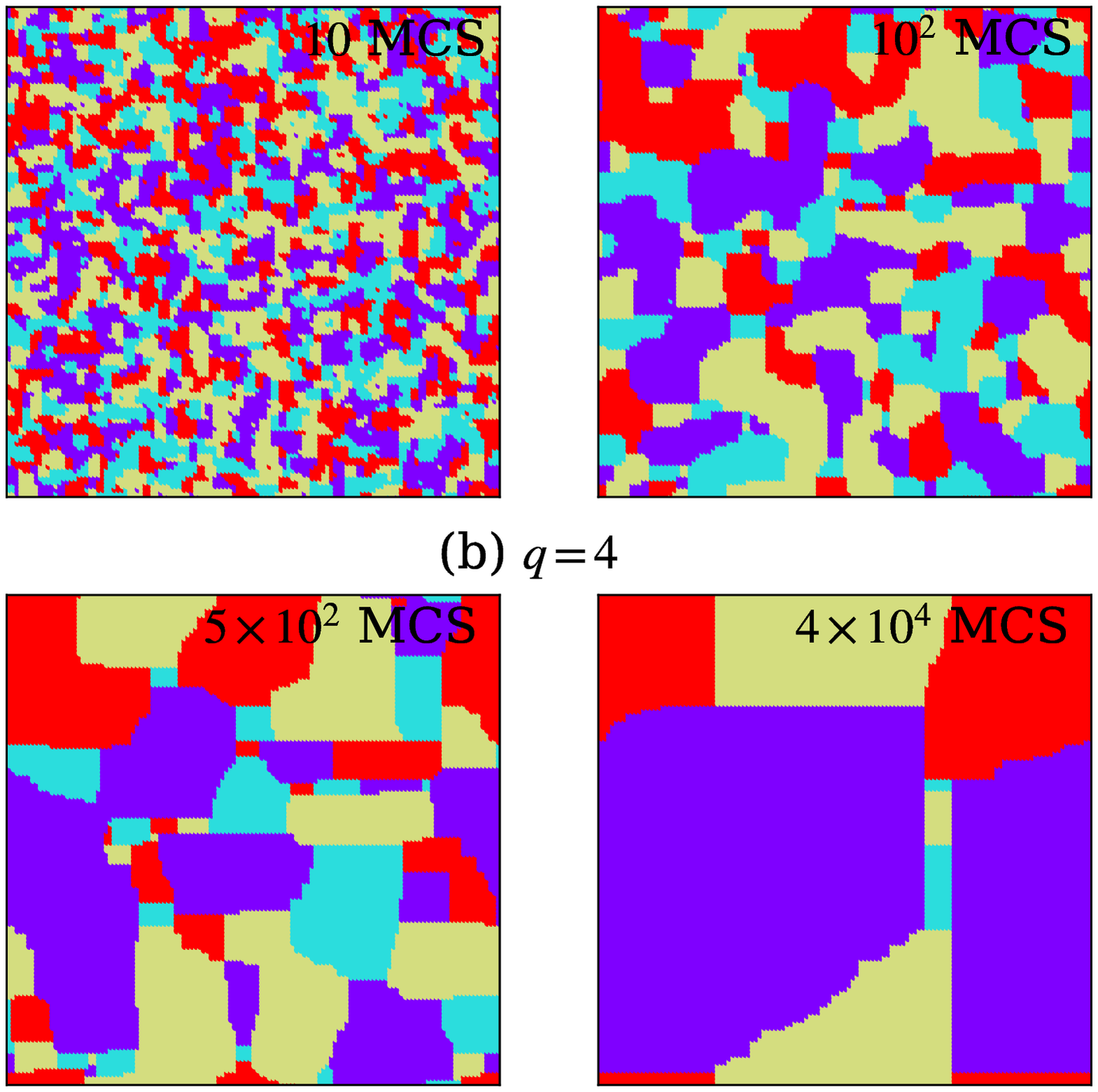}
\caption{\label{snap_T0} Representative snapshots illustrating time evolution during phase ordering of the Potts model with (a) $q=2$ and (b) $q=4$ at $T=0$ for a square lattice having $L=128$. Again, different colors correspond to different spin states or species.}
\end{figure*}
\par
From now onward, we will consider $\ell_1$ as the growing length scale to verify the expected scaling law for phase ordering in different cases. Additionally, we will investigate the behavior of domain lengths of the last ranked \textit{loser}, i.e., $\ell_q$. In Fig.\ \ref{length_diffq_T0.6} (a), we compare the time dependence of $\ell_1(t)$ for $2\le q \le 10$ using a square lattice of size $L=128$ at $T=0.6T_c$. Data for all $q$ follows the expected $\sim t^{1/2}$ behavior for an extended period until they are affected by finite-size effects. Importantly, none of them show any early-time corrections. The amplitude of growth decreases as $q$ increases. Same power-law growth with different amplitudes suggests that these growths can also be quantified via a universal finite-size scaling function with a non-universal metric factor \cite{majumder_Potts,Janke_CP}. We have checked that an analogous plot using the average domain length $\ell(t)$ cannot provide the same universal picture for such a small system size. 

\par
The universality is not restricted to the behavior of $\ell_1(t)$ but also gets reflected in the time dependence of $\ell_q(t)$ too, as shown in Fig.\ \ref{length_diffq_T0.6}(b). There, $\ell_q(t)$ for all values of $q$ exhibits a non monotonic behavior with first showing a growth followed by a decay. In the growth regime data for all $q$ are almost parallel to each other. Albeit, the growth is much slower than the LCA law. Then, for all $q$ there is a gradual crossover to the decay regime. There, at  long time, the data are again parallel to each other, and are consistent with the $\sim t^{-2}$ behavior. This strongly suggests the presence of a universal decay law 
\begin{equation}\label{decay_eqn}
 \ell_q(t)\sim t^{-z},
\end{equation}
where $z=2$ is the dynamical exponent. The theoretical origin of the above relation still needs to be investigated.

\par
We also check the system size dependence of the behavior of the domain lengths in Figs.\ \ref{length_diffL_T0.6}(a) and (b), respectively, for the \textit{winner} and \textit{loser}. In (a) the data for $\ell_1(t)$ from different $L$ follow each other until they enter the finite-size affected regime, and are consistent with the expected $\sim t^{1/2}$ behavior. In (b) too the data for smaller system sizes follow the larger ones up to a certain time, and then gradually start decaying. In the growth regime, data for all system sizes show a growth slower than the LCA behavior $\sim t^{1/2}$. Importantly, in the decay regime all data are consistent with $t^{-2}$ behavior, further numerically confirming the validity of Eq.\ \eqref{decay_eqn}.

\subsection{Phase ordering at $T=0$ in $d=2$ and $3$}\label{zeroT}
In the previous subsection we have established the usefulness of disentangling the growth of different spin states or species in terms of \textit{winner} and \textit{losers}. Here, we explore the same in investigating the special case of phase ordering in the $q$-state Potts model at $T=0$ in $d=2$ and also in $d=3$.

\subsubsection{$d=2$}\label{d=2_T0}
Phase ordering at $T=0$ in $d=2$ for $q=2$ Potts model that corresponds to the Ising model, is the least challenging of all where one typically observes an ideal representation of the associated scaling law. Of course, there could be occasional cases of dynamic freezing, but overall a major fraction of  performed simulations still reach the ground state in a finite time for finite lattices. A representative for time evolution of such a system using a square lattice of size $L=128$ is shown in Fig.\ \ref{snap_T0}(a). The configuration at the latest time ($t=3\times 10^4$ MCS) corresponds to a typical ground state at $T=0$, where only one of the spin states or species survives. 
\begin{figure}[t!]
\centering
\includegraphics*[width=0.45 \textwidth]{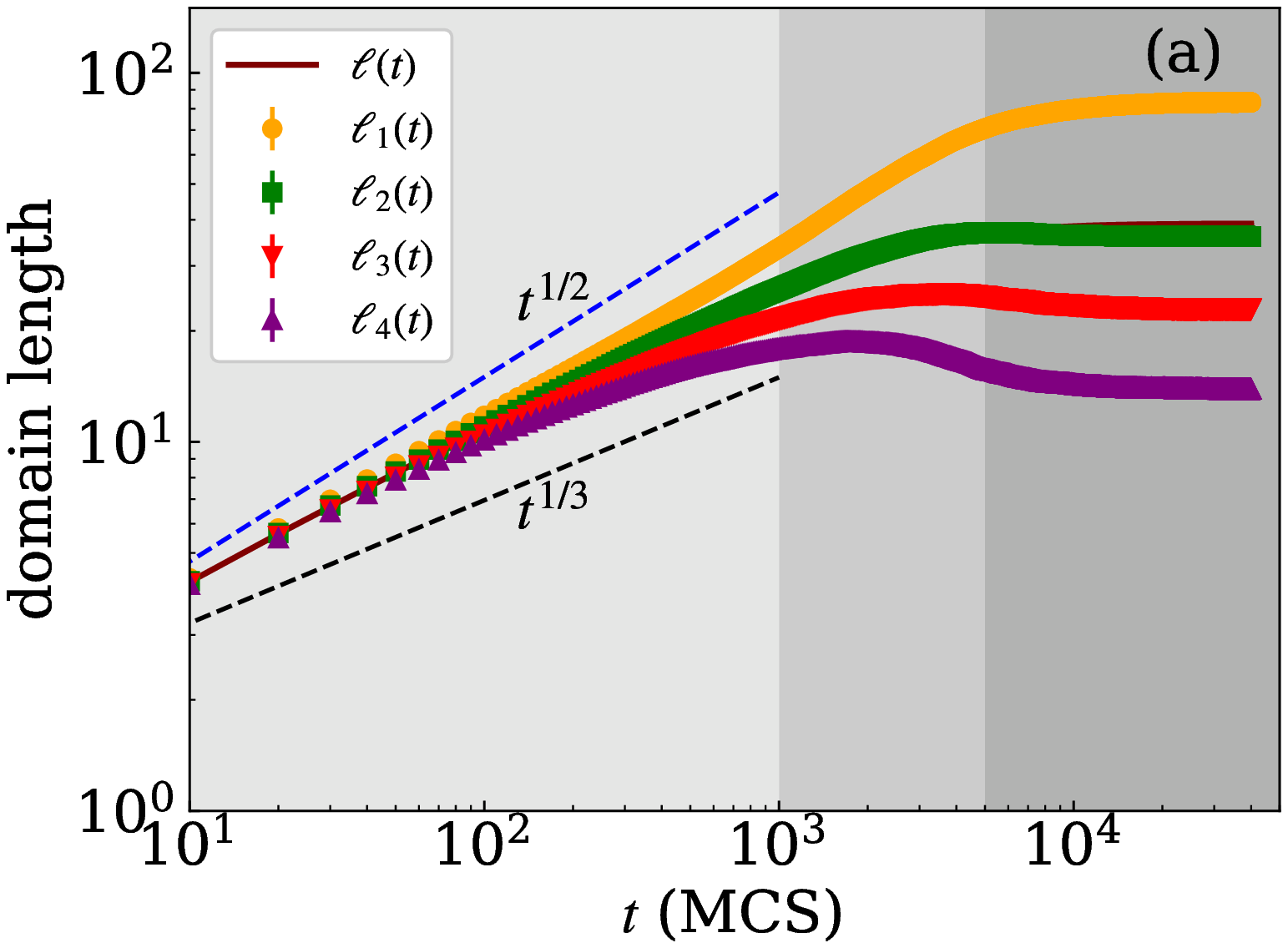}\\
\includegraphics*[width=0.47 \textwidth]{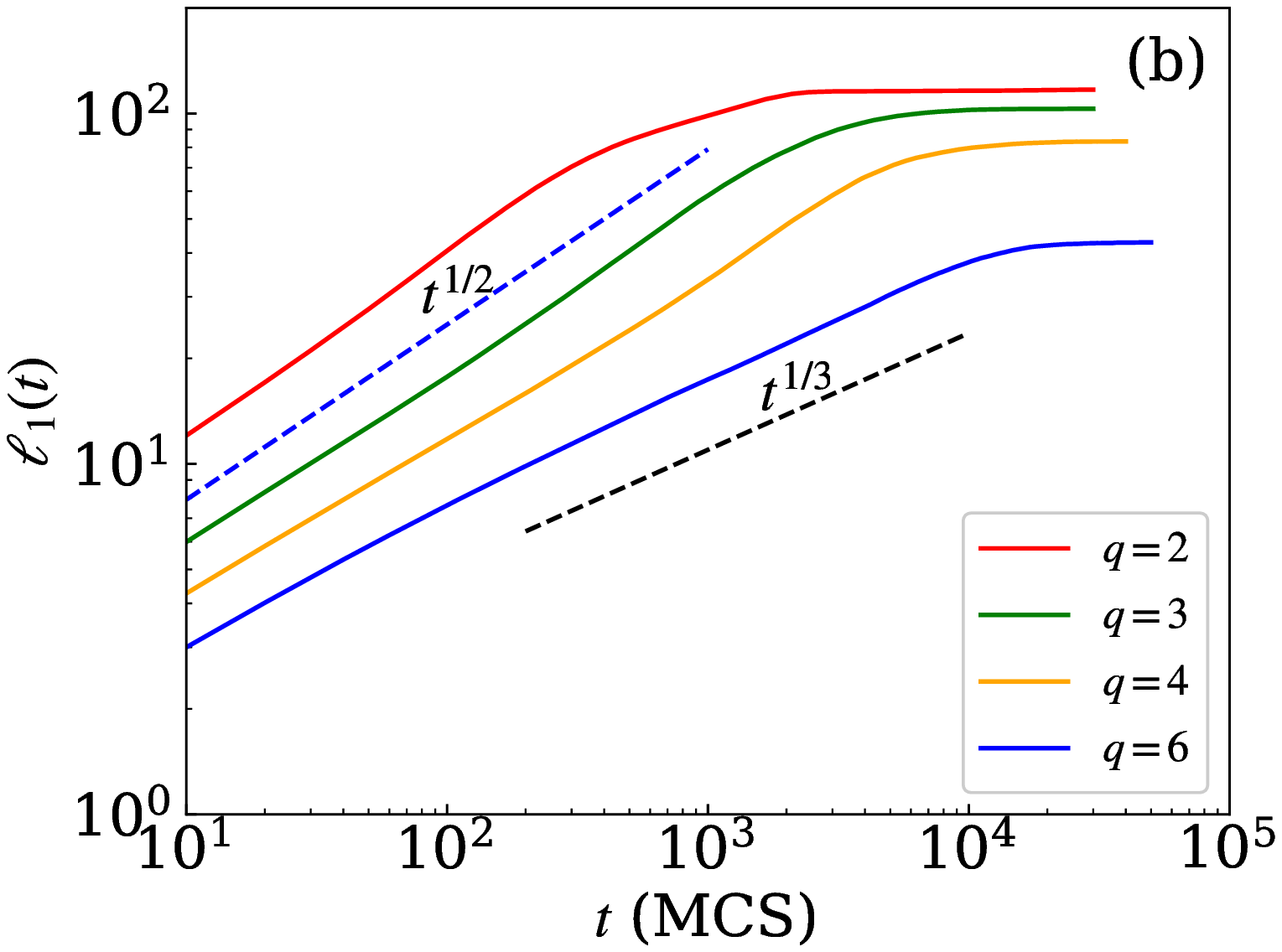}\\
\includegraphics*[width=0.47 \textwidth]{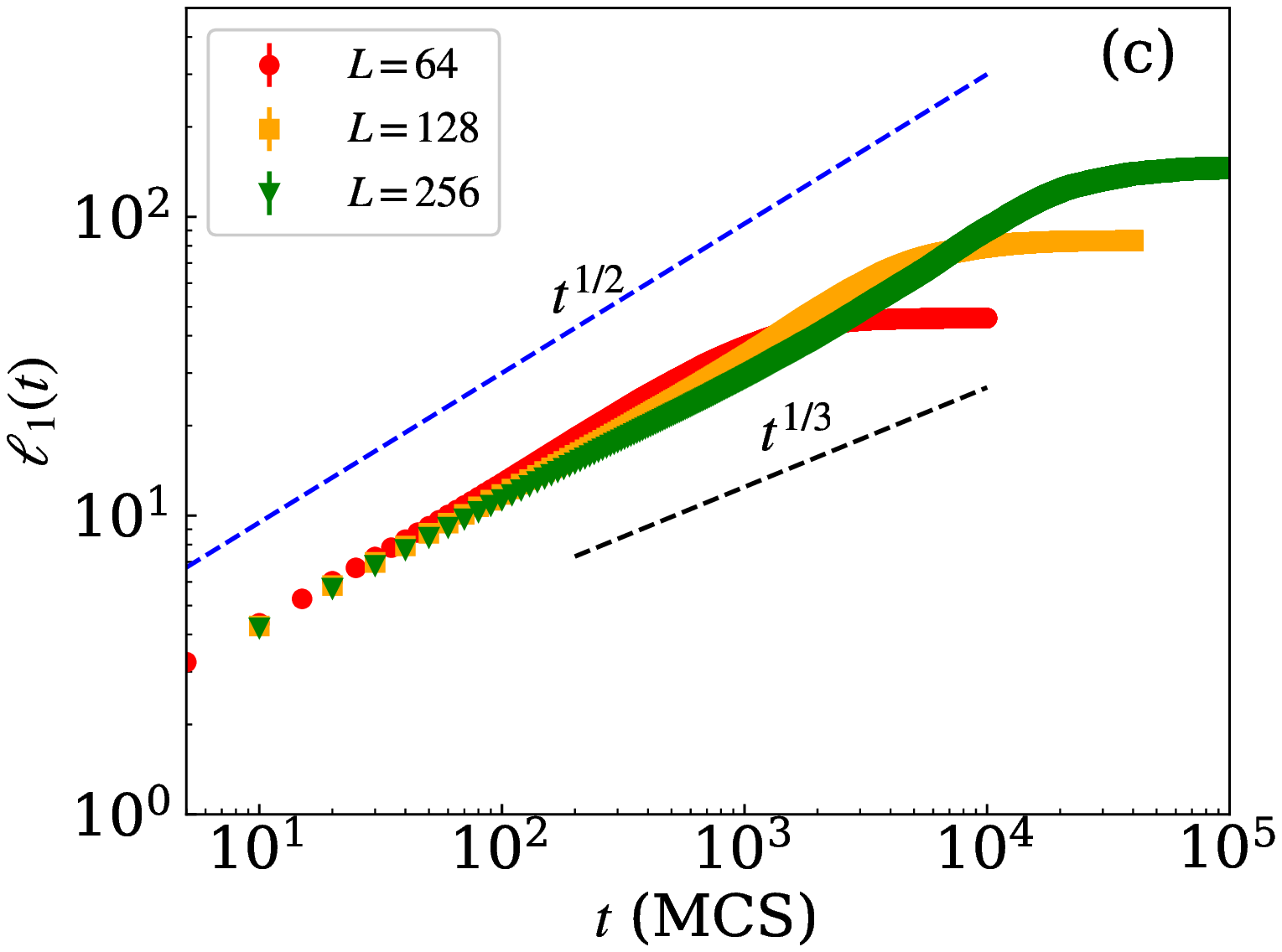}
\caption{\label{length_T0} (a) Time dependence of measured domain lengths of the \textit{winner} and \textit{losers} in $d=2$ for $q=4$ during phase ordering at $T=0$ using a square lattice with $L=128$. The traditionally measured average domain length $\ell(t)$ is also plotted. (b) Domain length $\ell_1$ of the \textit{winner} species as a function of time for different values of $q$ using the same square lattice. (c) System size dependence of $\ell_1(t)$ at $T=0$ for a fixed $q=4$. Dashed blue and black lines, respectively, represent the power-laws $\sim t^{1/2}$ and $\sim t^{1/3}$.}
\end{figure}

\par
For larger $q$, on the other hand, a significant fraction of the total simulations get stuck. This observation is in accordance with the known difficulty of reaching ground states for $q\ge d+1$ \cite{lifshitz1962,safran1981}.  It was also argued that for $q\ge 3$ in $d=2$ due the emergence of pinned configuration during the evolution, the system may get stuck in a disordered state leading to glassy dynamics \cite{petri2003,deOlivera2004,deOlivera2004_epl}. Relatively recently, from long time simulations of the model it was further confirmed that for $q\ge 2$ in $   
d=2$ the probability of reaching ground states is non-zero \cite{olejarz2013}. There, it was found that in the simulations where the system becomes static,  the configurations has perfectly flat interfaces. In Fig.\ \ref{snap_T0}(b) where we present the evolution of the $q=4$ Potts model, the snapshot at the latest time represent such a configuration with flat interfaces. Apart from these, we too observed existence of late time configurations which were evolving {\em ad infinitum}, as typically observed during phase ordering of the $d=3$ Ising model at $T=0$ \cite{olejarz2011,olejarz2011zero}. We have also performed simulation for other values of $q$ up to $q=10$. The overall picture remains the same although the probability of reaching the ground state within our maximum simulation time almost vanishes as $q$ increases, and episodes of observing pinned configuration become almost certain. 
\begin{figure}[t]
\centering
\includegraphics*[width=0.475 \textwidth]{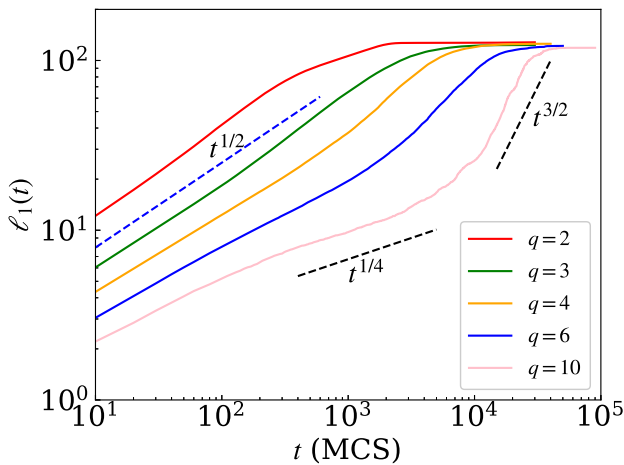}
\caption{\label{length_nonfreeze} Time dependence of the domain length $\ell_1$ obtained from simulations which did not freeze within the maximum simulation time,  at $T=0$ in $d=2$ for different $q$ using a system with $L=128$. Dashed blue and black lines represent possible power-law behaviors as indicated.}
\end{figure}
\par
Having discussed the difficulties of reaching ground states, we move on to study the domain-growth kinetics at $T=0$ by using our prescription of categorizing the different spin states or species as the \textit{winner} and \textit{losers}. Since for large $q$ a significant fraction of the simulations do not reach the ground state, it is not straightforward to decide the \textit{winner}. However, if one looks at the time evolution as in Fig.\ \ref{snap_T0}(b) for $q=4$, then one can still choose the purple spin state or species as the \textit{winner}, as it is the one that has been growing throughout, and has the largest size in the latest time. All the other spin states or species are thus considered to be \textit{losers}. Following this we estimate all possible domain lengths, and present the corresponding plots for $q=4$ in Fig.\ \ref{length_T0}(a). There, we have also shown the data for the average domain length $\ell(t)$ obtained by usual way of considering all spin states or species alike. The data for the \textit{losers} and $\ell(t)$ clearly show a growth much slower than the LCA growth. The data for $\ell_4$ in fact hints towards a power-law growth with an exponent $\alpha=1/3$. On the other hand, the data for the winner again is fairly consistent with the $\sim t^{1/2}$ growth. Fitting with the form 
\begin{equation}\label{fitting}
\ell_1(t)=At^{\alpha},
\end{equation} 
in the ranges $[10:3\times 10^3]$,  $[10^2:3\times 10^3]$, and $[2\time 10^2:3\times 10^3]$, respectively, yield $\alpha=0.453(1)$, $0.474(1)$, and $0.486(1)$. The slight deviation from $\alpha=1/2$ is due to the presence of pinned configurations that do not allow the system to reach its ground state within the time period up to which we run the simulations. Nevertheless, our target here is not to obtain the ground state, rather to apply our approach in quantifying the domain-growth kinetics with the data we obtain at $T=0$ after running the simulations for a finite time. Hence, we abstain ourselves in performing simulation for long times. Keeping in mind that the ground state is not reached always and looking at the long-time flat behavior of $\ell_4$ in Fig.\ \ref{length_T0}(b), here, we do not explore the validation of Eq.\ \eqref{decay_eqn}.

\par
In Fig.\ \ref{length_T0}(b) we present the domain-growth kinetics of the \textit{winner species} at $T=0$ for different values of $q$. For $q\le 4$ the growth is almost consistent with expected LCA growth with an exponent $\alpha=1/2$, until finite-size effects creep in. The data for $q=6$ shows significant deviation from the LCA right from early time, and at later times it is consistent with the $\sim t^{1/3}$ behavior. This is a virtue of the slow  dynamics for larger $q$ due to pinning effects which will be overcome eventually in the long time limit via thermally activated processes in the form of an \textit{avalanche} \cite{olejarz2013}. However, consistency of our data for $q=3$ and $4$ with the LCA law, suggest that the bound $q\le d+1$ is not a strict one. We present the system size dependency of kinetics of $\ell_1$ at $T=0$ in Fig.\ \ref{length_T0} (c), which show prefect consistent with the LCA growth for the smallest system size $L=64$. For the largest system, i.e., $L=256$, the data seem to be weakly deviating from the LCA growth, due to the slowly dynamics arising from pinning effects. This suggests that pinning effects increase with system size.

\par
It is evident from Fig.\ \ref{length_T0} that at $T=0$, the value where $\ell_1$ saturates at long $t$, does not approach the finite-size limiting value $\approx L$. This can be attributed to the same fact that at $T=0$ not all the simulations could attain the ground state with all sites having the same spin states or species. Since this effect increases as $q$ increases, one notices in Fig.\ \ref{length_T0} that the saturating value of $\ell_1$ decreases as a function of $q$. To ignore the effect of the frozen dynamics, we now separate out the simulation runs where the \textit{winner} species could achieve a length $\ell_1 \approx L$ until $t=10^5$ MCS, which is significant for a system size as small as $L=128$. The time dependence of $\ell_1$ obtained only from the non-freezed simulations are shown in Fig.\ \ref{length_nonfreeze} for different $q$.    Again data for $q\le 4$ are perfectly consistent with the LCA growth for an extended time period before encountering finite-size effects. In fact from a careful observation, one can appreciate the fact that the data are more consistent with LCA growth, in comparison with the data in Figs.\ \ref{length_T0}. Although not apparent in the plot, the increasing estimated error on $\ell_1$ as a function of $q$, is also a consequence of the fact that the fraction of non-freezed simulations decreases as $q$ increases.
\par
In Fig.\ \ref{length_nonfreeze}, one can also notice that for $q > 4$ at long times the growth of $\ell_1$ seems to be faster than the LCA growth. For the largest $q$, i.e., for $q=10$, apart from the long-time faster regime there exists an intermediate time period when the growth of $\ell_1$ is extremely slow, consistent with a $\sim t^{1/4}$ behavior. The slow phase correspond to the presence of pinned configurations which eventually breaks via an \textit{avalanche} leading to the long-time faster growth regime. For $q=10$ the faster growth is consistent with $\ell_1(t) \sim t^{3/2}$. Of course, since the pinning effect is smaller for small $q$, the slope of the faster-growth regime appears to be increasing as a function of $q$. A detail study of even larger $q$ values  is needed to extract 
any functional dependence and limiting value of the exponent as a function of $q$ during the \textit{avalanche} of pinned configurations.
\subsubsection{$d=3$}\label{d=3_T0}
After having an understanding of phase ordering in $d=2$ at zero temperature, we now move on to explore the other special case of domain-growth kinetics in $d=3$ at $T=0$. Here, we restrict ourselves to the unique case of the $q=2$ Potts model, i.e., the Ising model. For higher $q$ one needs to perform rejection free MC simulations which we will present elsewhere. 
\begin{figure}[t!]
\centering
\includegraphics*[width=0.48 \textwidth]{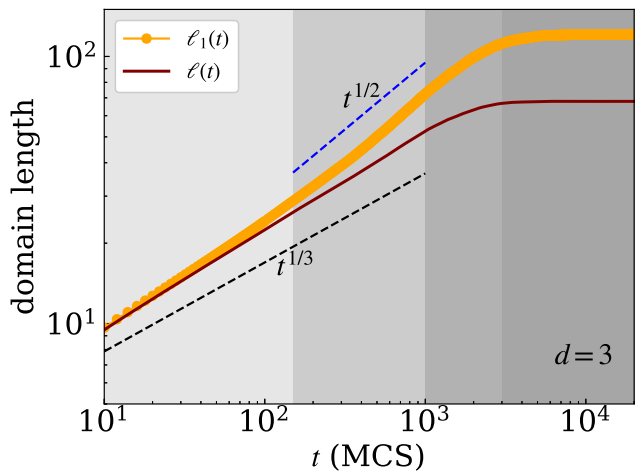}
\caption{\label{length_3d} Domain-growth kinetics in the $q=2$ Potts model, i.e., the Ising model, in terms of domain lengths $\ell$ and $\ell_1$, at $T=0$ in $d=3$. Data from simulations that did not get stuck are used. System used is a cubic lattice of length $L=128$. Dashed lines with different colors correspond to different power-law behaviors as indicated. Different shades are guides to the eyes to distinguish different regimes.}
\end{figure}
\par
The zero temperature phase ordering of the Ising model poses challenges because a major fraction of the simulations get stuck in local minima and thus one experiences frequent dynamic freezing. Hence, the measured time dependence of the average domain length deviates significantly from the expected LCA growth. In this regard, a dynamic crossover in the time dependence of the average domain length from a $\ell(t)\sim t^{1/3}$ at finite times to the asymptotic $\ell(t) \sim t^{1/2}$ behavior has been argued \cite{lipowski1999}, which was later confirmed numerically by considering data from simulations that did not get stuck \cite{vadakkayil2019}. This calls for the need of simulations of larger system sizes to access large length scale and time scales in order to realize the aforesaid crossover to the universal LCA growth. In all these previous studies, however, the growth of the domains of the \textit{winner} and \textit{loser} were not disentangled. In Fig.\ \ref{length_3d}, we present the kinetics of domain length of the \textit{winner} for the $q=2$ Potts model using a cubic lattice of $L=128$, at $T=0$ by considering only the simulations that evolved completely in the given maximum run time of our simulation, i.e., until $t=4\times 10^4$ MCS. There, we have also included the data for the average domain length $\ell(t)$ that shows the presence of a single scaling law which is slower than $\ell(t)\sim t^{1/2}$, albeit, faster than $\ell(t)\sim t^{1/3}$. A fitting with the form \eqref{fitting} in the range $[10:200]$ yields $\alpha=0.3703(3)$. On the other hand, $\ell_1(t)$ hints the presence of a crossover from a slower growth at early times to $\ell_1(t)\sim t^{1/2}$ behavior at long $t$. The early-time growth seems to be significantly faster than $\ell_1(t)\sim t^{1/3}$ behavior. Importantly, the crossover which was previously thought to be observed only at large length scale and time limit, is now realized much earlier in the behavior of $\ell_1$,  and for a much smaller system having linear dimension of $L=128$. 
\begin{figure}[t!]
\centering
\includegraphics*[width=0.48 \textwidth]{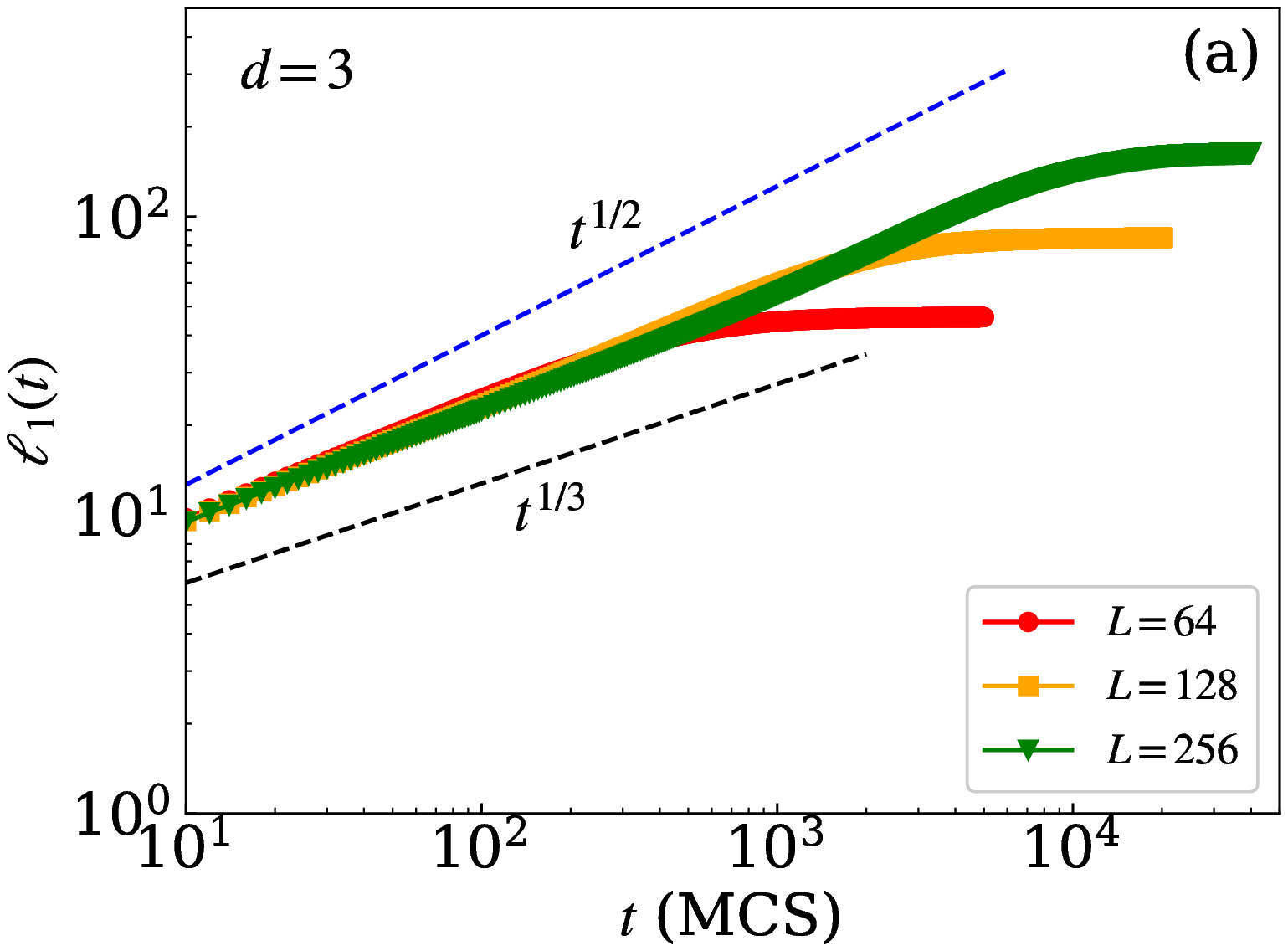}\\
\includegraphics*[width=0.48 \textwidth]{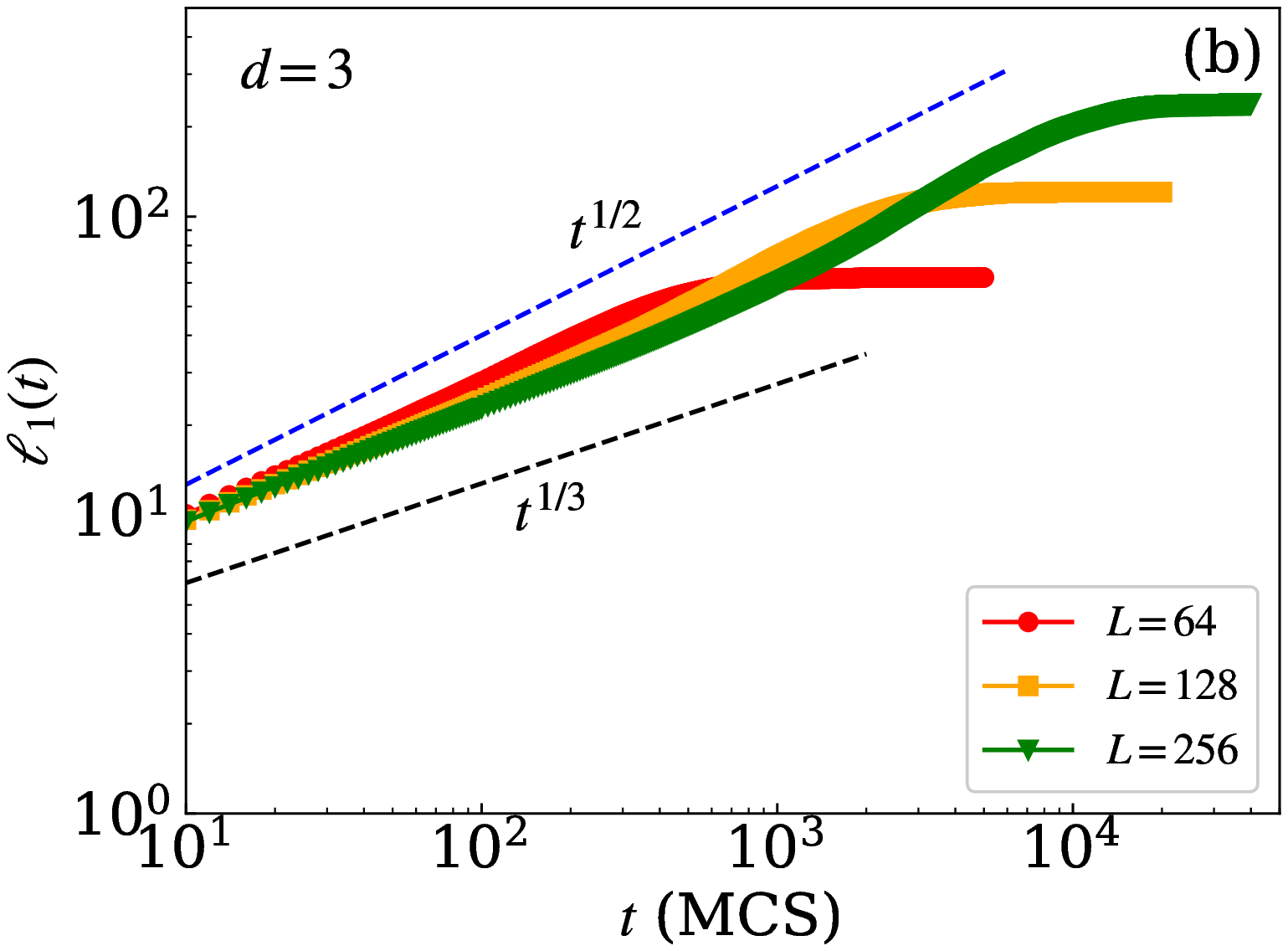}
\caption{\label{length_3d_diff_L} System size dependence of domain-growth kinetics of the \textit{winner} spin state or species for the $q=2$ Potts model, i.e., the Ising model, in $d=3$ at $T=0$. In (a) we show the data including the runs where the dynamics got frozen, whereas (b) shows the data excluding them. In both figures dashed lines with different colors correspond to different power-law behaviors, as indicated.}
\end{figure}
\par
In view of the results for $T=0$ in $d=2$, presented in the previous subsection, for the $d=3$ case too, one can expect the effect of pinned configurations. Thus, the crossover which is observed in Fig.\ \ref{length_3d} could supposedly be attributed to the pinning effect that stalls the dynamics for some time, and then overcome it via an \textit{avalanche}. To further substantiate this argument, in Fig.\ \ref{length_3d_diff_L} we present the data of $\ell_1(t)$ for different system for phase ordering systems having $q=2$ at $T=0$ in $d=3$. In Fig.\ \ref{length_3d_diff_L}(a) we show the plot $\ell_1(t)$ obtained by considering all the simulations that we ran, irrespective of the fact whether it got stuck within the maximum allowed run time. The data for different system sizes nicely overlap with each other, and are in agreement with the presence of a single scaling law which seems to be slower than $\ell(t) \sim t^{1/2}$, but faster than $\ell(t) \sim t^{1/3}$. A fitting of the data for $L=256$ using the form \eqref{fitting} in the range $[10:2\times 10^3]$ provides $\alpha=0.365(5)$. The corresponding plots  using data considering only the  simulations which could come out of the frozen dynamics within the given simulation time is presented in Fig.\ \ref{length_3d_diff_L}(b). There, the data for the smallest system size $L=64$ is perfectly consistent with $\ell_1(t)\sim t^{1/2}$ growth, with no indication of a crossover. While the absence of the crossover was expected as in previous studies using small system sizes, the realization of the LCA growth is contrastingly surprising. In Fig.\ \ref{length_3d_diff_L}(b), as the system size increases one can notice the emergence of a crossover from an early-time slower growth to a faster one. However, the  crossover time shifts as a function of $L$ implying its system size dependence.  This cannot be the case for a true crossover in the scaling. Thus, we infer that this is plausibly due to effect of pinned configurations that stalls the dynamics at finite times, and becomes more prominent as system size increases, similar to what we observed in the $d=2$ cases for $q\ge 4$. The system comes out of this pinned dynamics in the form of an \textit{avalanche} exhibiting a growth even faster than the LCA law. The data for the largest system size roughly indicate such a behavior in the post-crossover regime before it hits the finite-size limit. However, the exponent of the power-law seems to be smaller than $3/2$ which we obtained for the case of $q=10$ in $d=2$ in Fig.\ \ref{length_nonfreeze}.

\section{Conclusion}\label{conclusion}
In summary, we have presented results from phase-ordering dynamics of a multi-species system modeled via the $q$-state Potts model in space dimension $d=2$ and $3$. We have focused on disentangling the dynamics of different species on the basis of whether the species or spin state has survived as the dominating species in the final state. If the species is the majority one then we call it as the \textit{winner} or else  \textit{loser}. 
\par
In $d=2$ at a finite temperature $T=0.6T_c$, below the critical temperature $T_c$, we have observed that for $2 \le q \le 10$ the domain length $\ell_1$ of the  \textit{winner} shows nice agreement with the expected Lifshitz-Cahn-Allen (LCA) growth law $\ell_1(t) \sim t^{1/2}$ without any finite-time corrections for systems as small as a square lattice having a linear dimension of $L=128$. This defies the usual tradition of using large system sizes.  We have shown that such an observation with smaller system sizes cannot be realized if one uses the traditional average domain length $\ell(t)$ that is calculated by considering all spin states or species alike. We have also shown that the time dependence of the domain length of the \textit{loser} exhibits a rich dynamical behavior. At early times the domains of the \textit{loser} show a growth that is slower than the LCA growth. Followed by that it gradually starts decaying, and in the long-time limit exhibits a $\ell_q \sim t^{-z}$ behavior where $z=2$ is the dynamical exponent for nonconserved dynamics. To establish the universality of the above relation its validity needs to be verified for $d=3$. In the same spirit, it would be interesting to further verify it  for phase ordering in long-range interacting systems \cite{christiansen2019,janke2019,christiansen2021}.
\par
We have also explored the special case of phase ordering of the Potts model at $T=0$ in $d=2$. There, we have shown for $2 \le q \le 4$ that the time dependence of the domain length of the \textit{winner} is consistent the LCA growth, provided one has separated the simulations which did not get stuck due to frozen dynamics. The time dependence of $\ell_1(t)$ for large $q$ values show that at intermediate times the dynamics is stalled due to pinning. Subsequently, the system gets out of these pinned configuration which gets manifested in the form of an \textit{avalanche} exhibiting an ultra fast domain growth $\ell_1 \sim t^{3/2}$. In future, it would be worth to explore the scaling of the domain length in the \textit{avalanche} regime for larger $q$. For that one needs to implement the rejection free $n$-fold algorithm for faster computation \cite{BKL_1975}.
\par
In $d=3$, we have explored the special case of zero-temperature phase ordering for $q=2$, i.e., for the Ising model. We have shown that if one examines the time dependence of $\ell_1$, then the crossover in the growth from an early-time slower growth to a faster $\sim t^{1/2}$ growth can be realized in relatively smaller system size, e.g., in a cubic lattice of linear dimension $L=128$. Using data from different system sizes we have shown that these crossover in growth is system size dependent. That poses question about the true nature of the crossover. Rather, it indicates the presence of pinned dynamics which later is overcome in the form of an \textit{avalanche} as observed for higher $q$ in $d=2$. Nevertheless, in this regard, again a detailed study in $d=3$ for larger $q$ is needed, which we take up as a future endeavor. 

\acknowledgments
The work was funded by the Science and Engineering Research Board (SERB),
Govt. of India in the form of Ramanujan Fellowship under file no. RJF/2021/
000044. 


\end{document}